\documentclass[10pt, twocolumn, amsmath,amssymb,a4paper, pra]{revtex4}
\usepackage{theorem}
\usepackage[dvips]{graphicx}
\newcommand{\B}{\mathfrak{B}}

\newcommand{\Hi}{\mathcal{H}}

\newcommand{\Tr}{\mathrm{Tr}}

\newcommand{\lv}{\left \vert}
\newcommand{\rv}{\right \vert}
\newcommand{\la}{\left \langle}
\newcommand{\ra}{\right \rangle}
\newcommand{\ket}[1]{\lv #1 \ra}
\newcommand{\bra}[1]{\la #1 \rv}

\theorembodyfont{\rmfamily}
\newtheorem{Theorem}{{Theorem} }

\newtheorem{Definition}{{Definition} }
\newtheorem{Corollary}{{Corollary}}
\newtheorem{Lemma}{{Lemma}}
\newtheorem{Proof}{{Proof}}

\begin{document}

\title{Local copying and local
discrimination as a study for non-locality of a set}
\author{Masaki Owari$^{1,2}$, Masahito Hayashi$^{1,3}$}
\address{$^1${\it ERATO Quantum Computation and Information Project, JST, Tokyo 113-0033, Japan}\\
$^2${\it Department of Physics, The University of Tokyo, Tokyo 113-0033, Japan}
$^3${\it Superrobust Computation Project
Information Science and Technology Strategic Core (21st Century COE by MEXT)
Graduate School of Information Science and Technology
The University of Tokyo }}

\begin{abstract}
We focus on the non-locality concerning local copying and local
discrimination, especially for a set of orthogonal maximally
entangled states in prime dimensional systems, as a study of
non-locality of a set of states. As a result, for such a set, we
completely characterize deterministic local copiability and show
that local copying is more difficult than local discrimination.
From these result, we can conclude that lack algebraic symmetry
causes extra non-locality of a set.
\end{abstract}

\maketitle

\section{introduction}

Non-locality is one of the oldest topics in quantum physics, and
also is one of the most important topics in the newest fields,
``quantum information''. The history of non-locality started with
EPR's discussion of local realism in 30's \cite{epr}, and then, it
was followed Bell's formulation of local hidden variable theory
and Bell inequality in 60's \cite{bell}. In early 90's, the
development of quantum information shed a new light on this topic.
The theory of non-locality was reformulated as entanglement
theory, which is a useful formulation to treat entangled states as
resources of quantum communication, like teleportation,
dense-cording, key distribution, etc. \cite{teleportation, quantum
communication}. Mathematically speaking, the study of conventional
entanglement theory is the study of convertibility between
entangled states under locality restrictions for operations, (e.g.
LOCC (local operation and classical communication). Separable
operation, PPT (Positive Partial Transpose) operation \cite{LOCC,
BDSW, Morikoshi, JP, R01, relative entropy, convertibility}).

On the other hands, there are problems of non-locality which can
not be explained by one to one convertibility of states, that is,
``{\it Local Discrimination}''  (a problem to discriminate an
unknown states by only LOCC) and ``{\it Local Copying}'' (a
problem to copy an unknown states by only LOCC). The starting
point was a discover of a product basis which can not be perfectly
discriminated by LOCC \cite{non-locality} by Bennett {\it et. al.}
In \cite{non-locality}, they proposed a locally indistinguishable
product basis and regarded its impossibility for perfect
discrimination under LOCC as non-locality of it. After this work,
there have been many results of the local discrimination problems
\cite{WSHV00, VSPM01, CY02, C04, F04, VP03, GKRS02}. Recently, as
a similar problem to local discrimination, a new problem, ``{\it
local copying}'', was also raised \cite{ACP, GKR}, as a problem to
study a cloning of unknown entangled states under the LOCC
restriction with only minimum entanglement resource.

The study of Bennett {\it et. al.} suggests the new kind of
non-locality, {\it Non-locality of a set of states}. At first,
from analogy of the non-locality discussion in local
discrimination, we can expand the concept of non-locality as
follows. If the local (LOCC) restriction causes difficulty for a
task concerning a set of states, {\it e.g.} discrimination,
copying etc., then, we consider that this set has non-locality,
and regard the degree of this difficulty as {\it non-locality of
the set}. This concept of non-locality is not unnatural, since it
is consistent with the conventional entanglement theory because of
the following reasons. In entanglement theory, entanglement cost
\cite{BDSW} is one of the famous measures of entanglement, and can
be regarded as a kind of difficulty of a task, {\it i.e.} the
difficulty of entanglement dilution \cite{BDSW}. Moreover, if we
consider the task to approximate a given state by separable
states, we derive the relative entropy of entanglement
\cite{relative entropy} by measuring this difficulty in terms of
accuracy of the approximation, using relative entropy. These can
be regarded as the degrees of difficulty of tasks with the local
restrictions.

Indeed, local copying and local discrimination can be regarded as
tasks for a set of states with the local restriction, because
these problems are usually treated based on a set of candidates of
the unknown states. Hence, we can measure ``{\it Non-locality of a
set of states} by the degree of their difficulty. We should note
that this kinds of difficulty cannot be often characterized only
by entanglement of states of the given set. A typical example is
the impossibility of local discrimination of the product basis of
Bennett {et.al.} Actually, in addition to local discrimination and
local copying, the similar non-locality also appears commonly in
various different fields of quantum information, {\it e.g.}
quantum capacity, quantum estimation, etc. \cite{capacity,
estimation}. Since this type of non-locality does not depend only
on entanglement convertibility, {\it i.e.}, entanglement of each
states, we call it {\it Non-locality beyond individual
entanglement}

\begin{figure}[htbp]
\begin{center}
\includegraphics*[width = \linewidth ]{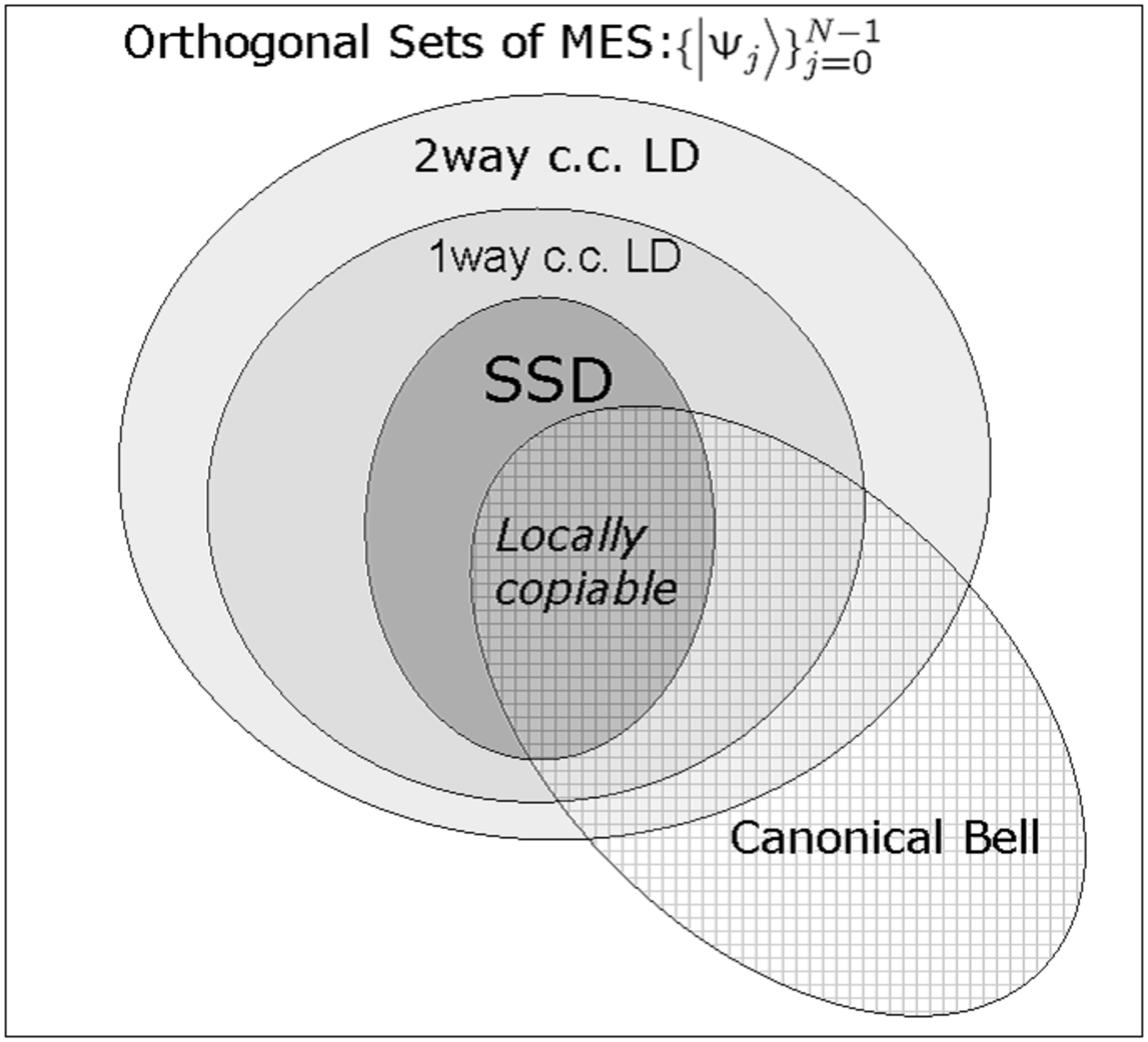}
\end{center}
\caption{The hierarchy of non-locality of sets of maximally
entangled states. \\ In this figure, LD, SSD, and c.c. mean
locally distinguishable,  Simultaneously Schmidt decomposable, and
classical communication, respectively.} \label{hierarchy}
\end{figure}

In this paper, as a study of non-locality beyond individual
entanglement, we focus on non-locality of a set of states by means
of local copying and local discrimination. Especially, we
concentrate ourselves on a set of orthogonal maximally entangled
states in a prime dimensional system for simplicity, and
investigate the relationship between their local copiability and
local distiniguishability. As a result, we completely characterize
the local copiability of such a set, that is, we prove that such a
set is locally copiable, if and only if it has canonical Bell form
and is simultaneous Schmidt decomposable. Using this result, we
prove the following two facts. First, the maximal size of locally
copiable sets is equal to the dimension of the local space as well
as the maximal size of local distinguishable sets. Second, we also
show that if such a set is locally copiable, then locally
distinguishable by one-way communication. Thus, in this case,
local copying is strictly more difficult than one-way local
discrimination. The relationship of local copiability and
distinguishability is summarized in FIG. \ref{hierarchy}.
From this relationship, we derive the conclusion related to the
non-locality of a set concerning local copying and local
discrimination: A simultaneous Schmidt decomposable state does not
have non-locality beyond individual entanglement concerning local
discrimination, since it is locally distinguishable. However, even
if a set is simultaneous Schmidt decomposable, if such a set does
not have a canonical Bell form, such a set still has non-locality
concerning local copying. Therefore, {\it lack of algebraic
symmetry causes extra non-locality of a set}.

Although we mainly concentrate ourselves on the aspect of local
copying and local discrimination as the study of non-locality in
this paper, local copying and local discrimination themselves are
worth to investigate as basic protocols of quantum information
processing with two-party. In the last part of this paper, we show
that there are many important relationships bewtween our local
copying protocol and the other quantum information protocols.
These results give many other interpritations for local copying.

This paper is organized as follows. In Section \ref{preliminary},
as a preparation of our analysis, we review a necessary and
sufficient condition for a locally copiable set as the preparation
of our analysis, which is the main result of the paper \cite{ACP}.
In Section \ref{local copying}, we give an example of a locally
copiable set of $D$ maximally entangled states, and then, prove
that, in a prime-dimensional local system, the above example is
the only case where local copying is possible. In Section
\ref{relation}, we discuss the relation between local copying and
LOCC discrimination by means of simultaneous Schmidt
decomposition. In Section \ref{application}, we present the other
protocols which are strongly related with our theory of local
copying, {\itshape i.e.}. channel copying, distillation protocol,
error correction, and quantum key distribution. And then, we
extend our results of local copying to these protocols. Finally,
we summarize and discuss our results in Section \ref{summary}.

\section{The local copying problem } \label{preliminary}
In this section, as a preparation of our analysis, we introduce
the formulation and the known results of local copying from
\cite{ACP}.

Many researchers treated approximated cloning, for example,
universal cloning \cite{universal cloning}, asymmetric cloning
\cite{asymmetric cloning}, tele-cloning \cite{tele-cloning}. This
is because the perfect cloning, {\it i.e.}, copying, is impossible
without prior knowledge ({\it no-cloning theorem })
\cite{no-cloning}. That is, the possibility of copying depends on
the prior knowledge, or, in other words, the set of candidates for
the unknown target state. If we know that the unknown state to be
copied is contained by the set of orthogonal states, which is
called the {\it copied set}, we can copy the given state. However,
if the system to be copied has an entangled structure, and if our
operation is restricted to local operations and classical
communications (LOCC)\cite{LOCC}, we cannot necessarily copy the
given quantum state with the above orthogonal assumption,
perfectly. Thus, it is interesting from both viewpoints of
entanglement theory and cloning theory to extend cloning problem
to the bipartite entangled setting. This is the original
motivation of cloning problems with LOCC restriction \cite{ACP,
GKR}.

Recently, F. Anselmi, {\it et. al.} \cite{ACP} focused on the perfect cloning of bipartite systems under the following assumptions;
\begin{enumerate}
\item Our operation is restricted to LOCC. \item It is known that
the unknown state to be copied is contained by the set of
orthogonal entangled states, (the copied set). \item An entangled
state of the same size is shared.
\end{enumerate}
They called this problem local copying, and they have
characterized copied sets which we can locally copied for special
cases. In the following, for simplicity, we say the set is locally
copiable if local copying is possible with the prior knowledge to
which the given state belongs. As is explained in Theorem
\ref{nscondition}, they showed that the possibility of local
copying can be reduced to the simultaneous transformation of
unitary operators. That is, they derived a necessary and
sufficient condition for a locally copiable set.

The problem of local copying can be arranged as the follows.
We assume two players at a long distance, {\itshape e.g.},
Alice and Bob in this protocol.
They have two quantum systems $\Hi _A$ and $\Hi _B$ each of which is also composed
 by the same  {\itshape two} $D$-dimensional systems,
{\itshape i.e.}, the systems $\Hi _A$ and $\Hi _B$
are described
by $\Hi _A = \Hi _1 \otimes \Hi _3$, $\Hi _B= \Hi _2 \otimes \Hi _4$.
In our problem, they try to copy an unknown state $\ket{\Psi}$ on the initial
system
$\Hi _1 \otimes \Hi _2$ to the target system $\Hi _3 \otimes \Hi _4$
with the prior knowledge that $\ket{\Psi}$ belongs to the copied set
$\{ \ket{\Psi _j } \}_{j=0}^{N-1}$.
Moreover,  we assume that they implement copying only by LOCC between them.
Since LOCC operations do not increase the entanglement of whole states,
they can copy no entangled state by LOCC without any entanglement resource.
Thus, we also assume that they share a blank entangled state $\ket{b}$ in target
systems $\Hi _3 \otimes \Hi _4$.
Therefore, a set of states $\{ \ket{\Psi _j } \}_{j=0}^{N-1}$
is called locally copiable with a blank state $\ket{b ^{3,4}} \in \Hi _3 \otimes
\Hi _4$,
if there exists
a LOCC operation $\Lambda$ on $\Hi _A \otimes \Hi _B$
which satisfies the following condition for all $ j = 0, \cdots , N-1$:
 \begin{eqnarray*}
& & \Lambda  (\ket{\Psi _j^{12}} \otimes \ket{b^{34}} \bra{\Psi _j^{12}} \otimes
 \bra{b^{34}} ) \\
 & = & \ket{\Psi _j^{12}} \otimes \ket{\Psi _j^{34}}
 \bra{\Psi _j^{12}} \otimes \bra{\Psi _j^{34}},
\end{eqnarray*}
 where we treat $\Hi _A = \Hi_1 \otimes \Hi _3$
 and $\Hi _B = \Hi _2
 \otimes \Hi_4$ as local spaces with respect to a LOCC operation $\Lambda$.
Since the local copying protocol is closely related to entanglement
catalysis \cite{ACP, JP}, that is well known open problem,
it is very hard to derive a necessary and sufficient condition for general
settings of local copying.
On the other hand, it is well known that no maximally entangled state
works as entanglement catalysis.
In this paper, to avoid the difficulty of entanglement catalysis,
we restrict our analysis to the case where all of $\ket{\Psi _j}$
are maximally entangled states
\footnote{The paper \cite{ACP} showed that
if a copied set $\{ \ket{\Psi _j} \}_{j=0}^{N-1}$ has at least one maximally
entangled state and is locally
copiable, then all of states $\ket{\Psi _j}$ in the copied set must be
maximally entangled.}.
Therefore, $\ket{\Psi _j} \in \Hi _1 \otimes \Hi _2$
can be represented by a unitary operation $U_j \in \Hi _1$ as
\begin{equation}
\ket{\Psi _j}=(U_j \otimes I) \ket{\Psi _0}.
\end{equation}

As the preparation of our paper, we shortly summarize
Anselmi, {\it et. al.}'s necessary and sufficient condition of local
copying as follows \cite{ACP}.
\begin{Theorem} \label{nscondition}
A set of maximally entangled states $\{ \ket{\Psi _j} \}_{j=0}^{N-1}$
is locally copiable, if and only if there exists a unitary operator $A$ on $\Hi _1
\otimes \Hi _3$ satisfying
\begin{equation} \label{nscondition1}
A(T_{jj^{'}} \otimes I)A^{\dagger}
=e^{i(\theta _j - \theta _{j^{'}})}(T_{jj^{'}} \otimes T_{jj^{'}}),
\end{equation}
where $T_{jj^{'}} = U_j U_{j^{'}}^{\dagger}$.
\end{Theorem}
Since each $\ket{\Psi _j}$ must be orthogonal, each $T_{jj^{'}}$ must satisfy
$\Tr ( T_{jj^{'}} ) = \delta _{jj^{'}}$.
Actually, we can derive the orthogonal conditions from Equation
(\ref{nscondition1}) by taking trace of them.
In the above theorem, the local copying operation $\Lambda $ is explicitly
represented as a local unitary transformation $A^{13} \otimes A^{*24}$.
Finally, they solved Equation (\ref{nscondition1}) for all $j, j^{'}$ in the case of $N=2$.
In this case,
there is only one independent equation, and  these conditions are reduced to
the condition $A(T \otimes
I)A^{\dagger} =T \otimes T$, where the phase factor $e^{i\theta _j}$ is
absorbed  by $T$.
The following theorem is the conclusion of their analysis of Equation
(\ref{nscondition1}) for $N =2$.
\begin{Theorem} \label{N=2}
There exists a unitary operator $A$ satisfying
\begin{equation} A(T
\otimes I) A^{\dagger} = T \otimes T,
\end{equation}
if and only if a unitary
operator $T$ satisfies the following two conditions:
\begin{enumerate}
\item The
spectrum of $T$ is the set of power of $M$th roots of unity, where $M$ is a
factor of $D$.

\item The distinct eigenvalues of $T$ have equal degeneracy.
\end{enumerate}
\end{Theorem}

Here, we should remark the number of maximally entangled states as the resource.
If we allow to use three entangled states as a resource, we can always locally copy any orthogonal set
of maximally entangled states by use of quantum teleportation \cite{teleportation},
(For the case when we share two entangled states as resources, see \cite{GKR}.)

\section{local copying of the maximally entangled states in prime-dimensional systems} \label{local copying}
In this section, we solve Equation (\ref{nscondition1}) and get the necessary
and sufficient condition  for all $N$
in the case of {\it prime}-$D$-dimensional local systems.
That is, the form of $T$ is completely determined.
As a consequence, we show that $D$ is the maximum size of a locally copiable set.

As the starting point of our analysis, we should remark that
Equation (\ref{nscondition1}) simultaneously presents $N^2$ matrix equations, but we may take care of only $N-1$ equations
$A (T _{j0} \otimes I) A^{\dagger} = e^{i \theta _j - i \theta _0} T _{j0} \otimes T _{j0} \quad (j = 0, \cdots , N-1)$.
This is because by multiplying the $j$ elements of the equation by the Hermitian conjugate of $j^{'}$ elements of
the same equation, we can recover Equation (\ref{nscondition1}).
Moreover, since $T _{j0}=U_j U_0^{\dagger} = U_j$ and the coefficient $e^{i \theta _j}$ is only related to the unphysical global phase factor,
we can treat only the following $N$ equations,
\begin{equation} \label{fundamental}
A (U _{j} \otimes I)A^{\dagger} = U _{j} \otimes U _{j} \quad (j = 0, \cdots , N-1).
\end{equation}
Note that $\ket{\Psi _j}$  is represented as $\ket{\Psi _j} = U_j \otimes I
\ket{\Psi _0}$.

\begin{figure}[htbp]
\begin{center}
\includegraphics*[width = \linewidth ]{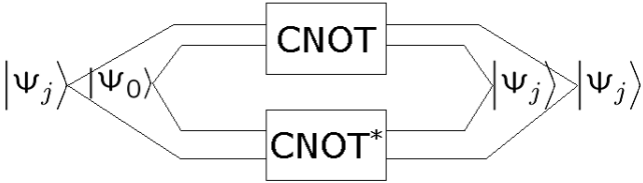}
\end{center}
\caption{The protocol of local copying}
\label{protocol-copy}
\end{figure}

At the first step, we construct an example of a locally copiable set of $D$ maximally entangled states.
\begin{Theorem}\label{sufficiency}
When the set of maximally entangled states $\{ \ket{\Psi _j} \}_{j=0}^{N-1}$ is defined by \begin{equation}
\ket{\Psi _j} = (U_j \otimes I) \ket{\Psi _j}
\end{equation}
and
\begin{equation} \label{definition}
U_j = \sum _{j=0}^{D-1} \omega ^{jk} \ket{k} \bra{k}, \end{equation}
where $\{ \ket{k} \}_{k=0}^{D-1}$ is an orthonormal basis of the $\Hi _1$, then
the set $\{ \ket{\Psi _j} \}_{j=0}^{N-1}$ can be locally copied.
\end{Theorem}
\begin{Proof}
We define the unitary operator $A$ by
\begin{equation} \label{repA}
A = {\rm CNOT}\stackrel{\rm def}{=} \sum _{a,b} \ket{a \ominus b} \ket{b} \bra{a} \bra{b},
\end{equation}
where $\rm CNOT$ is an extension of Control-NOT gate represented in $\{ \ket{k} \}_{k=0}^{D-1}$for $D$ dimensional systems.
Then, we can easily verify Equation (\ref{fundamental}) as
\begin{eqnarray*}
& & A(U_j \otimes I)A^{\dagger} \\
& = & {\rm CNOT} (U_j \otimes I) {\rm CNOT}^{\dagger} \\
& = & \sum \ket{a_1 \ominus b_1} \ket{b_1} \bra{a_1} \bra{b_1} (\omega ^{j a_3} \ket{a_3} \bra{a_3}) \\
& & \qquad \ket{a_2} \ket{b_2} \bra{a_2 \ominus b_2} \bra{b_2} \\ & = & \sum _{a_1, b_1}
\omega ^{ja_1} \ket{a_1 \ominus b_1} \bra{a_1 \ominus b_1} \otimes \ket{b_1} \bra{b_1} \\ & = & \sum _{c, b_1}
\omega ^{j (b_1 \oplus c) } \ket{c} \bra{c} \otimes \ket{b_1} \bra{b_1} \\ & = & U_j \otimes U _j,
\end{eqnarray*}
where we set $c = a_1 \ominus b_1$.
Therefore, Theorem \ref{nscondition} guarantees that the set $\{ \ket{\Psi _j}
\}_{j = 0}^{D-1}$ can be locally copied.
\hfill $\square$
\end{Proof}
This protocol of local copying used is the above proof is written as FIG \ref{protocol-copy}.
Here, we should remark that $U _1$ is the generalized Pauli's $Z$ operator which is one of
the generators of the Weyl-Heisenberg Group, and another $U _j$ is the $j$th power of $U_1 = Z$.
Hence, in the case of non-prime-dimensional local systems,
the spectrum of $U _j$ is different from that of $U _1$
if $j$ is a non-trivial factor of $D$.

Moreover, the property of Weyl-Heisenberg Group not only guarantees that the above example
satisfies  (\ref{fundamental}), but also is essential for the condition
(\ref{fundamental}).
That is, as is proved below, any locally copiable set of maximally entangled states
is restricted exclusively to the above example.
Therefore, our main theorem can be written down as follows.
\begin{Theorem} \label{main}
In prime-dimensional local systems, the set of maximally entangled states $\{ U_j \otimes I \ket{\Psi _0}
\}_{j=0}^{N-1}$
can be locally copied if and only if there exist an orthonormal basis $\{ \ket{a }\}_{a=0}^{D-1}$ and
a set of integers $\{n _j \}_{j=0}^{N-1}$ such that the unitary $U _j$ can be written as
\begin{equation} \label{nsrep}
U_j = \sum _{k=0}^{D-1} \omega ^{n_j k} \ket{k} \bra{k},
\end{equation}
where $\omega$ is the $D$th root of unity.
\end{Theorem}
Since the size of the set $\{ U_j \}$ is $D$, $D$, that is equal
to the dimension of local space, is the maximum size of a locally
copiable set of maximally entangled states with prime-dimensional
local systems. In comparison with the case without LOCC
restriction, this is actually the square root.

The proof of Theorem \ref{main} is as follows.
\begin{Proof}
(If part)
We have already proven that $\{ U_j \otimes I \ket{\Psi _j} \}_{j=0}^{D-1}$ can be copied by LOCC in
Theorem \ref{sufficiency}.
Therefore, the subset of them can be trivially copied by LOCC.

(Only if part)
Assume that a unitary operator $A$ satisfies the condition (\ref{fundamental}) for all $j$.

By applying Theorem \ref{N=2}, we can choose an orthonormal basis $\{ \ket{a} \}_{a=0}^{D-1}$ such that
\begin{equation} \label{expression}
U_1 = \sum _{a=0}^{D-1} \omega ^a \ket{a} \bra{a},
\end{equation}
where $\omega$ is $D$th root of unity.
Moreover, Equation (\ref{fundamental}) implies that the unitary $A$ should transform the subspace $\ket{a} \otimes \Hi$ to subspace $span \{ \ket{k} \otimes \ket{l} \}_{k \oplus l = a}$.
That is, $A$ is expressed as
\begin{equation} \label{repA}
A = \sum _{a,b,c} \xi _{b,c}^{a} \ket{a \ominus c} \ket{c} \bra{a} \bra{b},
\end{equation}
where $\xi _{b,c}^{a}$ is a unitary matrix for $b, c$ for the same $a$, that is,
$\sum _{c = 0}^{D-1} \xi _{b,c}^{a} \overline{\xi}_{b^{'} ,c}^{a} = \delta _{b,b^{'}} $
 and $\sum _{b = 0}^{D-1} \xi _{b,c}^{a} \overline{\xi}_{b ,c^{'}}^{a} = \delta _{c,c^{'}}$.
Thus, based on the basis $\{ \ket{a} \}_{a=0}^{D-1}$, Equation (\ref{fundamental}) for all
$a_1, a_2, b_1, b_2$ is written down as
\begin{equation} \label{elementeq}
\bra{a_1}\bra{b_1} A(U_j \otimes 1) A^{\dagger} \ket{a_2}\ket{b_2} =\bra{a_1} U_j \ket{a_2} \bra{b_1} U_j \ket{b_2}.
\end{equation}
Therefore, substituting Equation (\ref{repA}) to Equation (\ref{elementeq}) for any integer $j$, we obtain

\begin{eqnarray} \label{matrixeq}
\sum _{b=0}^{D-1} \xi _{b,b_1}^{a_1 \oplus b_1} \overline{\xi} _{b, b_2}^{a_2 \oplus b_2}
\bra{a_1 \oplus b_1} U_j \ket{a_2 \oplus b_2} \nonumber \\ = \bra{a_1} U_j \ket{a _2} \bra{b_1}U_j \ket{b_2},
\end{eqnarray}
for all $a _1, a_2, b_1$ and $b_2$.

To see that $U _1 $ and $U _j$ can be simultaneous orthogonalized, we need to prove the following lemma.
\begin{Lemma} \label{lemma}
A non-zero $D \times D$ matrix $U_{ab}$ satisfies the following equation,
\begin{equation} \label{eqlemma1}
\Xi _{b_1 b_2}^{a_1 \oplus b_1 , a_2 \oplus b_2} U_{a_1 \oplus b_1 \ a_2 \oplus b_2}
=U_{a_1 a_2} U_{b_1 b_2},
\end{equation}
where $\Xi _{b_1 b_2} ^{c c} = \delta _{b_1 b_2}$ and all indices have their value between $0$ and $D - 1$, then $U_{ab}$ is a diagonal matrix.
\end{Lemma}
\begin{Proof}
See Appendix A
\end{Proof}


We apply this Lemma \ref{lemma} to the case when $U_{ab} = \bra{a} U_j \ket{b}$ and $\Xi _{b_1 \ b_2}^{a_1 \ a_2} =
\sum _{b = 0}^{D-1} \xi _{b b_1}^{a_1} \overline{\xi} _{b b_2}^{a_2}$.
Then, this lemma shows that $U_j$ is orthogonal in the eigenbasis of $U_1$, therefore all
unitaries $\{U_j \}_{j=0}^{N-1}$ are orthogonalized.
Then, we can get the form of $\{ U_j \}_{j=0}^{N-1}$ explicitly as follows.
From the diagonal element of (\ref{matrixeq}), we derive
\begin{equation} \label{group}
\bra{a \oplus b} U_j \ket{a \oplus b} = \bra{a} U_j \ket{a} \bra{b} U_j \ket{b}.
\end{equation}
Since $\{ \ket{a} \}_{a=0}^{D-1}$ is also an eigenbasis of $U_j$, we can express $U _j $ as
\begin{equation} \label{unitary}
U_j = \sum _{a = 0}^{D-1} \omega^{P_j (a)} \ket{a} \bra{a},
\end{equation}
where $P_j (a) $ is a bijection from $\{ a \}_{a=0}^{D-1}$ to themselves.
Then, Equation (\ref{group}) guarantees that $P_j (a)$ is a self-isomorphism of the cyclic group
$\{ a \}_{a=0}^{D-1}$.
Since a self-isomorphism of a cyclic group is identified by the image of the generator,
 we derive the formula (\ref{nsrep}) with $P_j(1) = n_j$.
\hfill $\square$
\end{Proof}

We have solved the LOCC copying problem only for a {\it
prime}-dimensional local space. In the case of a
non-prime-dimensional local space, our proof of the ``{\it only if
part}'' can be done in the same way. However, the ``{\it if
part}'' is extended straightly only for the case in which the set
$\{ U_j \}_{j=0}^{N-1}$ contains at least one unitary whose
eigenvalues are generated by the $D$th root of unity. In this
case, the proof is the following. By the same procedure of the
prime-dimensional case, we obtain Equation (\ref{matrixeq}). Then,
Lemma \ref{lemma} implies that all $U_j$ can be diagonalized, and
also implies Equation (\ref{group}) for all $U_j$. By writing
$U_j$ as (\ref{unitary}), we get the equation $P_j (a \oplus b) =
P _j(a) \oplus P _j(b)$ and, so, $P _j(a) = a P _j(1)$. Hence,
Theorem \ref{N=2} guarantees the same representation of $U_j$ as
(\ref{nsrep}). Therefore, we can solve the problem of local
copying in non-prime-dimensional local spaces only in this special
case as the direct extension of Theorem \ref{main}. On the other
hand, if eigenvalues of all $U_j$ are degenerate, our proof of
``{\it if part}'' does not hold.

\section{relation with LOCC copying and LOCC discrimination} \label{relation}
If we have no LOCC restriction, the possibility of the
deterministic copying is equivalent with that of the perfect
distinguishability. However, we can easily see that under the
restriction of LOCC, this relation is non trivial at all. As we
have already mentioned in the introduction, these two problems
share the common feature, that is, their difficulty can be regard
as a non-locality of a set, and
this non-locality can not be explained only by entanglement
convertibility.
Therefore, the study of this relationship is really important to
understand the non-locality of a set.
In this section, we compare the locally distinguishability and the
locally copiability for a set of orthogonal maximally entangled
states. Thus, by introducing Simultaneous Schmidt decomposition,
we show the relationship between these two non-locality.

At first, we remind the definition of a locally distinguishable
set, and then mention several known and new results of locally
distinguishability. A set of states $\{ \ket{\Psi _j}
\}_{j=0}^{N-1}$ is called two-way (one-way) classical communication (c.c.)
locally distinguishable, if there exists
a POVM $\{ M_j \}_{j=0}^{N-1}$ which can be constructed by two-way (one-way)
LOCC
and also satisfies the following conditions:
\begin{equation}
\forall i,j ,\quad \bra{\Psi _i} M_j \ket{\Psi _i} = \delta _{ij}.
\end{equation}
In order to compare LOCC copying and LOCC discrimination, we should take care of the following point:
We assume an extra maximally entangled state
only in the LOCC copying case.
This is because
LOCC copying of a set of maximally entangled states is trivially impossible without a blank entangled state.
This fact is contrary to LOCC discrimination since LOCC discrimination requires
sharing no maximally entangled state.

In the previous section, we have already proved that $D$ is the maximum
size of locally copiable set of maximally entangled states. In the
case of local discrimination, we can also prove that $D$ is the
maximum size of a locally distinguishable set of maximally
entangled states. This statement was proved by the paper
\cite{GKRS02} only when the set of maximally entangled states $\{
\ket{\Psi _j} \}_{j = 0}^{N-1}$ consists of canonical form Bell
states, where a canonical form Bell state $\ket{\Psi _{nm}}$ is
defined as
\begin{eqnarray*}
\ket{\Psi _{nm}} & \stackrel{\rm def}{=} & Z^n X^m \otimes I \ket{\Psi _{00}} \\
\ket{\Psi_{00}} & \stackrel{\rm def}{=} & \sum _{k=0}^{d-1} \ket{k}\otimes \ket{k}\\
X & \stackrel{\rm def}{=}& \sum _{k=1}^d \ket{k}\bra{k \oplus 1}.
\end{eqnarray*}
Such a set is a special case of a set of maximally entangled states.
Here, we give a simple proof of this statement for a general set of maximally entangled states by the same technique as \cite{Tsuda}.
\begin{Theorem} \label{dika}
If an orthogonal set of maximally entangled states $\{ \ket{\Psi _j} \} _{j=0}^{N-1}$ is locally distinguishable,
then $N \le D$.
\end{Theorem}
\begin{Proof}
Suppose that $\{ M_j \}_{j =0}^{N-1}$ is a separable POVM which distinguishes $\{ \ket{\Psi _j} \}_{j=0}^{N-1}$,
then they can be decomposed as
$M _i = \sum _{k=0}^{L} p_{ik} \ket{\psi _k}\bra{\psi _k} \otimes \ket{\phi _k} \bra{\phi _k}$, where $p_{ik}$
is a positive coefficient.
Then, we can derive an upper bound of $\bra{\Psi _j} M_i \ket{\Psi _j}$ as follows,
\begin{eqnarray*}
\bra{\Psi _j} M_i \ket{\Psi _j} & = & \sum _{k=0}^{L} p_{ik} \bra{\Psi _j} \ket{\psi _k}\bra{\psi _k} \otimes
                                     \ket{\phi _k} \bra{\phi} \ket{\Psi _j} \\
                                & \le & \sum _{k=0}^{L} p_{ik} \bra{\psi _k} ( \frac{1}{D} I) \ket{\psi _k} \\
                                & \le & \frac{\Tr M _i}{D},
\end{eqnarray*}
where the first inequality comes from the montonisity of the fidelity under partial trace operations
concerning the system $B$.
Since $\bra{\Psi _j} M_j \ket{\Psi _j} = 1$, we have $1 \le \Tr ( M _j ) / D$.
Finally, taking the summation of the inequality for $j$, we obtain $N \le D^2/D=D$.
\hfill $\square$
\end{Proof}
Therefore, in this case, the maximal size of both locally copiable
and locally distinguishable sets is equal to the dimension of the
local space.

When we consider the relationship between local discrimination and local copying of a set of maximally entangled states,
it is quite useful to introduce ``\textit{Simultaneous Schmidt Decomposition}'' \cite{HH04}.
A set of states $\{ \ket{\Psi _{\alpha}} \} _{\alpha \in \Gamma} \subset \Hi _1 \otimes \Hi _2$ is
called simultaneously Schmidt decomposable, if they can be written down as
\begin{equation}
\ket{\Psi _{\alpha}} = \sum _{k=0}^{d-1} b_k^{( \alpha )} \ket{e_k} \ket{f_k},
\end{equation}
where $\Gamma$ is a parameter set, $\{ \ket{e_k} \}_{k=0}^{d-1}$ and $\{ \ket{f_k} \}_{k=0}^{d-1}$ are
orthonormal bases of local spaces
(simultaneous Schmidt basis) and $b_k^{(\alpha )}$ is a {\itshape complex number} coefficient.
Actually, for a set of orthogonal maximally entangled states,
simultaneous Schmidt decomposability is a sufficient condition for
one-way local distinguishability
and a necessary condition for local copiability of it.
Moreover, simultaneous Schmidt decomposability is not a
necessary and sufficient condition for the both cases.
Therefore, a family of locally copiable sets of maximally entangled states is
strictly included by
a family of one-way locally distinguishable sets of maximally entangled
states.
In the following, we prove this relationship.

First, we explain the relationship between local discrimination and simultaneous Schmidt decomposition which has
been already obtained by the paper \cite{VSPM01}.
If an unknown state $\ket{\Psi _{\alpha}} \in \Hi _A \otimes \Hi _B$ is in
a simultaneously Schmidt decomposable set of states $\{ \ket{ \Psi _{\alpha}} \} _{\alpha \in \Gamma}$, such a state can be
transformed to a single local space $\Hi _A$ or $\Hi _B$ by LOCC.
Rigorously speaking, there exists a LOCC $\Lambda$ on $\Hi _A \otimes \Hi _{B_1 B_2}$ which transforms
$\ket{\Psi _{\alpha}^{AB_1}} \otimes \ket{0^{B_2}}$ to $\sigma^A \otimes \ket{\Psi _{\alpha}^{B_1 B_2}}$
for all $\alpha \in \Gamma$, and also exists a LOCC $\Lambda ^{'}$on $\Hi _{A_1} \otimes \Hi _{A_2} \otimes \Hi _B$
which transforms
$\ket{0^{A_1}}  \otimes \ket{\Psi _{\alpha}^{A_2B}} $ to $\ket{\Psi _{\alpha}^{A_1 A_2}} \otimes \sigma^B  $
for all $\alpha \in \Gamma$.
Indeed, this LOCC transformation can be written down as the following Kraus representation \cite{VSPM01}:
\begin{eqnarray*}
&\rho \mapsto
\sum _{k=0}^{d-1} F_k
\rho
F_k^*,
\end{eqnarray*}
where
\begin{eqnarray*}
F_k
  & \stackrel{\rm def}{=} &
(I_A \otimes CNOT)
(U_k \otimes I_{A,B_2})
(P_k \otimes I_{B_1,B_2}) \\
P_k & \stackrel{\rm def}{=} & 1/D(\sum _{i } \omega ^{ki} \ket{e_i})(\sum _{l } \omega ^{kl} \bra{e_l}) \\
U_k & \stackrel{\rm def}{=} & \sum _{i} \omega ^{ki} \ket{f_i} \bra{f_i} \\
CNOT & \stackrel{\rm def}{=} & \sum _{kl} \ket{e_k} \otimes \ket{f_{k \oplus l}} \bra{f_k} \otimes \bra{l}.
\end{eqnarray*}
\begin{figure}[htbp] \label{SSD protocol}
\begin{center}
\includegraphics[width=\linewidth]{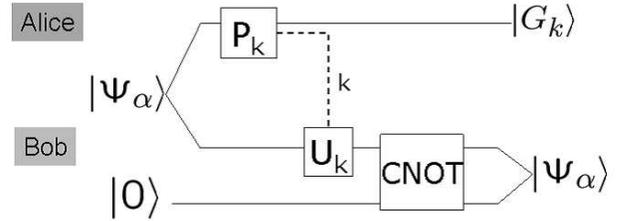}
\end{center}
\caption{a set of simultaneous Schmidt decomposable states can be send to Bob's space by LOCC }
\label{picture-discrimination}
\end{figure}
In the above formula, both $\{ \ket{e_k} \}_{k=0}^{D-1}$ and $\{ \ket{f_l} \}_{k=0}^{D-1}$ are
simultaneous Schmidt bases of $\{ \ket{\Psi _{\alpha}} \}_{\alpha \in {\Gamma}}$,
and $\{ \ket{l} \}_{l=0}^{D-1}$ is the standard computational basis.
This protocol can be written as FIG \ref{picture-discrimination}, where $\ket{G _{k}}$ is a garbage state with no information.
Using the above protocol,
if a set $\{ \ket{\Psi _{\alpha} }\}$ is
simultaneously Schmidt decomposable,
there exists a one-way-LOCC POVM $M'=\{M'_i\}$
for a given arbitrary POVM $M=\{M_i\}$ such that
\begin{eqnarray*}
\bra{\Psi _{\alpha}}  M_i \ket{\Psi _{\alpha}}
=
\bra{\Psi _{\alpha}}  M'_i \ket{\Psi _{\alpha}} ,\quad
\forall i, \forall \alpha.
\end{eqnarray*}
That is, any POVM can be essentially realized by 1way LOCC.
Therefore, ``\textit{simultaneously Schmidt decomposable set of
orthogonal maximally entangled states is one-way locally
distinguishable.}'' On the other hand, the set of orthogonal
maximally entangled states which is not simultaneously Schmidt
decomposable was found by the paper \cite{GKRS02}. Thus, a family
of simultaneously Schmidt decomposable sets of maximally entangled
states is strictly included by a family of locally distinguishable
sets of maximally entangled states.

On the other hand, the relationship between simultaneous Schmidt decomposability and
local copiability can be derived as the following theorem.
\begin{Theorem} \label{main}
In prime-dimensional local systems, an orthogonal set of maximally entangled states $\{ \ket{\Psi _j} \}_{j=0}^{N-1}$
is locally copiable,
if and only if it is a  simultaneously Schmidt decomposable
subset of canonical form Bell states under the same local unitary operation.
\end{Theorem}
\begin{Proof}
We can easily see the ``{\itshape only if}'' part of the above statement from Theorem \ref{main}.
The ``{\itshape if}'' part can be showed as follows.
The paper \cite{HH04} shows that the states $\ket{\Psi _{n_{\alpha} m_{\alpha}}} (\alpha = 1,2, \cdots , l)$
are simultaneously Schmidt decomposable, if and only if there exist integers $p$, $q$ and $r$
($p \neq 0$ or $q \neq 0$) satisfying $p n_{\alpha} \oplus q m_{\alpha} = r$ for all $\alpha$.
Since the ring ${\mathbb Z}_p$ is a field in the prime number $p$ case, the
above condition is reduced to the existence of $f$ and $g$ such that $m_{\alpha} =f n_{\alpha} + g$.
Then, we get
\begin{eqnarray}
\ket{\Psi _{n_{\alpha} m_{\alpha}}} & = & \ket{\Psi _{n_{\alpha} (f n_{\alpha} +g)}} \nonumber \\
                                    & = & C_{\alpha} (Z X^f)^{n_{\alpha}} X^g \otimes I \ket{\Psi _{00}}. \label{eqssd}
\end{eqnarray}
Since $Z X^f$ is unitary equivalent to $Z$ \cite{F04}, the state $\ket{\Psi _{n_{\alpha} m_{\alpha}}} $ is locally unitary
equivalent with $U_j \otimes I \ket{\Psi _0}$ in Theorem \ref{main}.
\hfill $\square$
\end{Proof}
We add a remark here. Under the assumption of simultaneous Schmidt
decomposition, a set has cannonical Bell form, if and only if the
set of corresponding unitary operators is a cyclic group, that is,
the group with only one generator. Therefore, we can rephrase this
necessary and sufficient condition as the follows, the set is
simultaneous Schmidt decomposable and satisfies the following
condition by a renumbering,
\begin{eqnarray} \label{cyclic}
U_1^{D} = I, \ U_k = \overbrace{U_1  \cdots U_1}^k.
\end{eqnarray}

Finally, we derive FIG.\ref{hierarchy},
and, therefore, for maximally entangled states,
a family of locally copiable sets is strictly included by
a family of simultaneously Schmidt decomposable sets.
In other words, local copying is more difficult than local discrimination.

At this last part of this section, we discuss our main results in
FIG \ref{hierarchy}, in the view point of non-locality beyond
individual entanglement.

In the case of bipartite pure states, all information of a
bipartite state $\ket{\Psi } = \sum _{i=0}^{D-1} \lambda _i
\ket{e_i} \otimes \ket{f _i}$ can be separated to two parts, that
is, Schmidt coefficients $\lambda _i$ and Schmidt basis $\{
\ket{e_i}, \ket{f_i} \}_{i=0}^{D-1}$, where $\lambda _i \ge 0$.
Because of local unitary equivalence, Schmidt
coefficients completely determine entanglement convertibility
\cite{LOCC}. Therefore, conversely, we can regard the non-locality
coming from interrelationship among Schmidt basis
as non-locality
purely beyond individual entanglement. In the following discussion,
we try to separate non-locality which depends on Schmidt coefficients
and Schmidt basis.

At first, since all sets in FIG. \ref{hierarchy} have the same Schmidt
coefficients, the structure of non-locality in FIG.\ref{hierarchy}
is determined only by the interrelationship of
Schmidt basis, and the effect of Schmidt coefficients do not
appear directly in this figure.
On the other hand,
since Schmidt basis do not concern
the definition of the maximal
sizes of local distinguishable and copiable sets,
the maximal sizes depend only on Schmidt coefficients.
Therefore, Schmidt coefficients
may affect only the maximal size of local distinguishable and
copiable sets.

The interrelationship of Schmidt basis like
is determined by the unitary
operator $U = \sum _{i=0}^{D-1} \ket{e_i} \bra{f_i}$.
In FIG.\ref{hierarchy}, the two properties of the interrelationship of
Schmidt basis, that is, such unitary operators, are
related to non-locality of a set. That is, simultaneous Schmidt
decomposability and canonical Bell form seems to reduce
non-locality of a set.
For simultaneous Schmidt
decomposable sets, we can explain their lack of non-locality as
follows.
As we well know, in the case of pure bipartite states,
one person can always apply the local operation which causes the same transformation
for a given state as another person's local operation causes (Lo-Popescu's theorem \cite{Lo-Popescu}).
The simple structure of entanglement
convertibility originates in the above symmetry
between local systems.
This symmetry is caused by the existence of Schmidt
decomposition. Similarly, in the case of local discrimination,
the protocol FIG. \ref{SSD protocol} seems to utilize this kind of
symmetry between local systems. Therefore, the existence of
simultaneous Schmidt decomposable basis gives the symmetry between
the local systems, and this fact may decreases the non-locality of the
sets of states.

In the case of cannonical Bell form, the interesting fact is that
this algebraic property is related to local copiability, and not
to local distinguishability.
As we have already seen, Since a simultaneous Schmidt decomposable set can be
transformed to a single local space by LOCC,
we can use any global discrimination protocols to such a set by only LOCC.
Therefore, concerning local discrimination, the sets of
simultaneous Schmidt decomposable states seem not to possess any non-locality
which originates in interrelationship between their Schmidt basis.
However, if such a set does not have a cannonical Bell form,
it is not locally copiable.
That is, a set has extra non-locality beyond individual entanglement
concerning local copying, if it has no algebraic structure given in (\ref{cyclic})  ,
even if it is simultaneous Schmidt decomposable.
Finally, we can conclude that, in the view point of problems
of non-locality beyond individual entanglement,
the above algebraic non-locality is most remarkable difference
between local copying and local discrimination.


\section{application to channel, distillation, and error correction} \label{application}
So far, we have treated local copying mainly in the context of non-locality
of a set. On the other hand, since local copying
of maximally entangled states is one of fundamental two-party protocol,
this problem itself is worth to investigate.
In this last section, we apply our results,
especially Theorem \ref{main} for different contexts from local
copying, and give several other interpretations for our results,
like channel copying, entanglement distillation, error correction,
and QKD. Thus, these many connections imply the fruitfulness of
local copying problem as a fundamental two-party protocol.
Moreover, seeing local copying problem from these
various points of view, we may also derive some clue which help us to
construct further development of understanding of non-locality
beyond entanglement convertibility.

\subsection{channel copying} \label{cc}
In section \ref{local copying}, in the analysis of local copying,
we treated not directly maximally entangled states, but unitary
operators which represent the maximally entangled states based on
a some standard maximally entangled states. This method is a kind
of operator algebraic method, or physically speaking equal to
Heisenberg picture. Therefore, we can interpret our results as
directly the results for these unitary operators themselves. Then, as a
result, we can derive the result for the problem of
``{\it unitary channel copying }''.

Here, we consider a problem ``{\it channel copying }'', that is, a problem
in which we ask a question as follows,
in the case we do not have complete description of a channel,
``Can we simulate two copies of the channel by using the channel only once?''.
As we will see in the following discussion, channel copying
with help of one-way classical communication (c.c.) is equivalent to local
 copying of
corresponding entangled states with help of one-way c.c..
The problem setting of channel copying can be written down as follows,
\begin{Definition}
We call that a set of channel $\{ \Lambda _{i} \}_{i = 1}^N ; \B
(\Hi _{A1}) \rightarrow \B (\Hi _{B1})$ is copiable with one way
c.c. and a blank channel $\Lambda _b$, if for all $i$, there
exists sets of Kraus's operators $\{ A_k \}_{k=1}^K \subset \B
(\Hi _{A1} \otimes \Hi _{A2})$, $\{ B^k_l \}_{l=1}^L \subset
\B(\Hi _{B1} \otimes \Hi _{B2} )$ such that $\sum _{k=1}^K A
_k^{\dagger} A_k =I_{A}, \sum _{l=1}^L B ^{k \dagger}_l B^k _l =
I_B$ for all $l$, and for all $i$ and $\rho$ on $\Hi _{A1} \otimes
\Hi _{A2}$,
\begin{equation} \label{channel copy}
\sum _{k l} B^k_l [\Lambda _i \otimes \Lambda _b (A_k \rho A^{\dagger}_k) ] B^{k \dagger}_l = \Lambda _i \otimes \Lambda _i (\rho).
\end{equation}
\end{Definition}
The meaning of the above definition can be sketched as the FIG
\ref{channel copy}, that is, by an encoding operation $\{ A_k
\}_{k=1}^K$ and a decoding operation $\{ B^k_l \}_{l=1}^L$,
$\Lambda _i$ with a blank channel (may be noisy) $\Lambda _b$
works as $\Lambda _i \otimes \Lambda _i$.
\begin{figure}[htbp]
\begin{center}
\includegraphics[width=\linewidth]{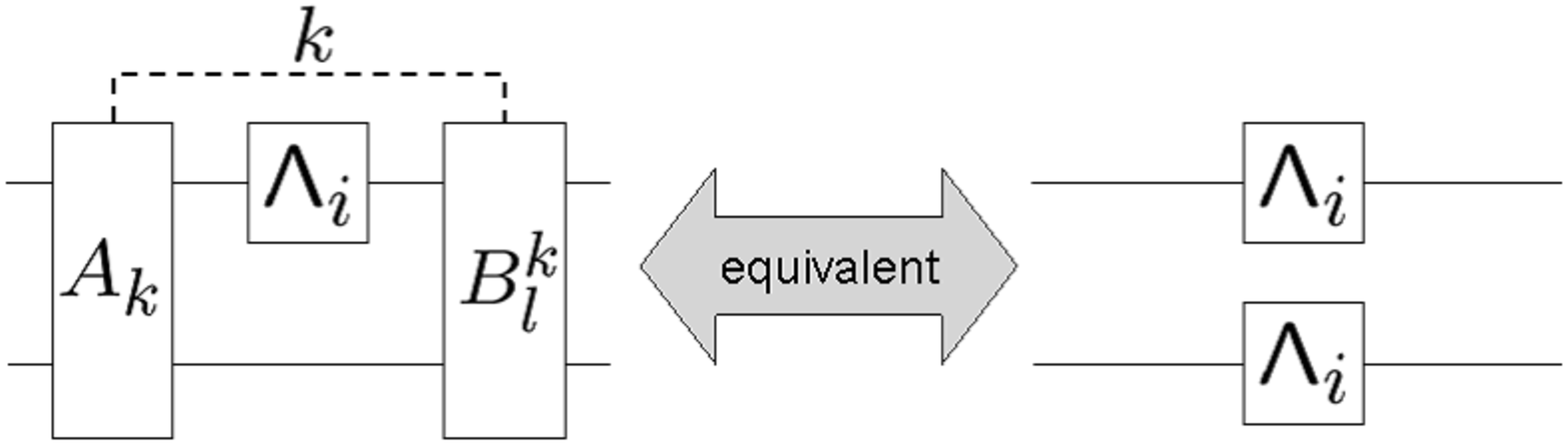}
\end{center}
\caption{Definition of channel copying with one way classical
communication \\
By suitable encoding $\{ A_k \}$ and decoding $\{ B^k_l \}$
operations, each $\Lambda _i$
works as $\Lambda _i \otimes \Lambda _i$.} \label{channel copy}
\end{figure}

Then, we can easily show  that the channel copying problem with
one way c.c. is exactly same as the local copying of
corresponding entangled states with one way c.c..
\begin{Theorem} \label{channel copy theorem}
A set of channel $\{ \Lambda _i \}_{i=0}^{N-1} $ is copiable with
one way c.c. and a blank channel $\Lambda _b$, if and only if a
set of entangled states $\{ \Lambda _i \otimes I (\ket{\Psi}
\bra{\Psi }) \}_{i=N}$ is locally copiable with one way c.c. and a
blank states $\Lambda _b \otimes I ( \ket{\Psi} \bra{\Psi})$,
where $\ket{\Psi}$ is an arbitrarily fixed maximally entangled
states.
\end{Theorem}
\begin{Proof}
Suppose $\{ \Lambda _i \}_{i=0}^{N-1} $ is copiable with one way c.c. and a blank channel $\Lambda _b$.
Consider four systems $\Hi _1 \otimes \Hi _2 \otimes \Hi _3 \otimes \Hi _4$,
and prepare two copies of maximally entangled states $\ket{\Psi} \bra{\Psi}$ on $\Hi _1 \otimes \Hi _3$ and $\Hi _2 \otimes \Hi _4$,
respectively.
Then, by applying channel copying protocol for $\Hi _1 \otimes \Hi _2$, we derive the following calculations.
\begin{eqnarray*}
&\quad & \sum _{kl} B_l^{k12} \otimes I^{34} [ \Lambda _i^1 \otimes \Lambda _b^2 \otimes I ^{34}  (A_k^{12} \otimes I^{34} \ket{\Psi ^{13}}\bra{\Psi ^{13}} \\
&\quad & \quad \otimes \ket{\Psi ^{24}} \bra{\Psi ^{24}}  A_k^{\dagger 12} \otimes I^{34})] B_l^{k \dagger 12} \otimes I^{34} \\
&=& \sum _{kl} B_l^{k12} \otimes I^{34} [ \Lambda _i^1 \otimes \Lambda _b^2 \otimes I ^{34} ( I^{12} \otimes A_k^{t34}  \ket{\Psi ^{13}}\bra{\Psi ^{13}} \\
&\quad& \quad \otimes \ket{\Psi ^{24}} \bra{\Psi ^{24}} I^{12} \otimes A_k^{* 34} )] B_l^{k \dagger 12} \otimes I^{34}\\
&=& \sum _{kl} B_l^{k12} \otimes A_k^{t34}  [(\Lambda _i^1 \otimes I^3 ( \ket{\Psi ^{13}}\bra{\Psi ^{13}}) \\
&\quad& \quad \otimes (\Lambda _b^2 \otimes I^4 \ket{\Psi ^{24}}\bra{\Psi ^{24}}) ]B_l^{k \dagger 12} \otimes A_k^{*  34} \\
&=& \Lambda _i^1 \otimes I^3 (\ket{\Psi ^{13}} \bra{\Psi ^{13}} ) \otimes \Lambda _i^2 \otimes I^4 (\ket{\Psi ^{24}} \bra{\Psi ^{24}}),
\end{eqnarray*}
where the last equality come from Eq. (\ref{channel copy}).
Therefore, $\Lambda _i \otimes I (\ket{\Psi } \bra{\Psi } )$ is
locally copiable with one-way classical communication. We can also
easily check the opposite direction of the proof. \hfill $\square$
\end{Proof}
The correspondence between channels $\Lambda$ and entangled states
$\Lambda \otimes I (\ket{\Psi} \bra{\Psi})$ is called
Jamilkowski's isomorphism \cite{Jamilkowski}. The above theorem
shows that the channel copying problems can be always identified
to corresponding local copying problems of entangled states in the
case of one-way classical communication. On the other hands, since
not all states can be written down as $\Lambda \otimes I
(\ket{\Psi } \bra{\Psi})$ for some maximally entangled state
$\ket{\Psi }$, not all local discrimination problems can be
considered as a channel copying problem.

Choosing all $\Lambda _i$ as
unitary channels, we derive maximally entangled states for
corresponding entangled states. Therefore, our results in Section
\ref{local copying} and \ref{relation} give also results for
unitary channel copying as follows.
\begin{Corollary}
In prime-dimensional systems, a set of Unitary channels $\{
\Lambda _i \}_{i=0}^{N-1}$ where $\Lambda _i (\rho ) = U_i \rho
U_i^{\dagger}$ is copiable with blank noiseless channel $\Lambda
_b = I$ and one way classical communication, if and only if $\{ U
_i \}_{i=0}^{N-1}$ is a simultaneous diagonalizable subset of
Weil-Heisenberg (Generalized Pauli) Group.
\end{Corollary}
\begin{Proof}
We can easily see from Theorem \ref{main} and \ref{channel copy
theorem}. \hfill $\square$
\end{Proof}

The above correspondence between local copying and channel copying
come from the fact that interrelationship of Schmidt basis can be
always represented by actions of unitary operators. Hence, we can
regard this mathematical correspondence as the one method to
represent the non-locality of sets of Schmidt basis in the
operational form for the corresponding unitary operations.

\subsection{Local copying of mixed states and Entanglement distillation}
As the next example of applications of our results, we consider
local copying of mixed states and entanglement distillation.
Although we have only considered the local copying of pure states
so far, we apply our protocol for mixed states in forward and also
backward directions in this subsection.

At first, if we apply our local copying protocol FIG.\ref{protocol-copy} of
sets of maximally entangled states $\{ \ket{\Psi _i} \} _{i=0}^{N-1}$
to the mixed states
$\rho _{in}= \sum _{i=0}^{N-1} p _i \ket{\Psi _i^{12}} \bra{\Psi _i^{12}}$,
then, we derive states
$\rho _{out} = \sum _{i=0}^{N-1} p _i \ket{\Psi _i^{12}} \bra{\Psi _i^{12} }
 \otimes  \ket{\Psi _i^{34}} \bra{\Psi _i^{34}}$ as a result.
Since the output state $\rho _{in}$ is equivalent to
$\Tr _{12} \rho _{out}$, and also to $\Tr _{34} \rho _{out}$,
we seems to succeed in copying these mixed states.
However, if we take account of the optimality of entanglement resource,
there would be a more efficient protocol for each choice of individual probability distribution $\{ p _i \}$. Therefore,
if we try to understand the correspondence between local copying of a set of
maximally entangled states and local copying of mixed states,
it would be better to define the local copiability of mixed states
for a set of states as well as that of pure states.
That is, we define local copiability as follows,
``a sets of mixed states $\{ \rho _{\xi }^{12} \} _{\xi \in \Xi }$ is called
locally copiable with blank states $\sigma ^{34}$,
 if there exists a LOCC $\Lambda $ such that for all
$\xi \in \Xi$,  $\Tr _{12} \Lambda (\rho _{\xi } \otimes \sigma ^{34} ) =
\Tr _{34} \Lambda (\rho _{\xi } \otimes \sigma ^{34}) = \rho _{\xi}^{12}$''.
Then, easily, we can translate the local copiability of a set of pure states
to that of mixed states as follows,
``a set of maximally entangled states
$\{ \ket{\Psi _i} \} _{i=0}^{N-1}$ is locally copiable,
if and only if a set of mixed states $\{ \rho _{ p } \} _{p \in P }$ is locally
 copiable with $\ket{\Psi _0}$, where
$\rho _p \stackrel{\rm def}{=} \sum _{i=0}^{N-1} p_i \ket{\Psi _i}
\bra{\Psi _i}$ and $P$ is a set of all probability distribution on
a set $\{ 0, \cdots, N-1 \}$.''

Since our local copying protocol
consists of local unitary operations, we can also consider the opposite
direction of our protocol. This inverse of local copying protocol
is actually entanglement distillation protocol by local unitary.
As we can see in FIG \ref{protocol-copy}, the inverse of our
protocol transforms $\ket{\Psi _i} \otimes \ket{\Psi _i} $ to
$\ket{\Psi _i} \otimes \ket{\Psi _0}$ by a local unitary
operation. Therefore, if we consider mixed states like
\begin{equation} \label{maximally correlated}
\rho = \sum _{ij} a _{ij} \ket{\Psi _i} \otimes \ket{\Psi _i}
\bra{\Psi _j} \otimes \bra{\Psi _j},
\end{equation}
and apply our local copying protocol, where $\{ \ket{\Psi _i}
\}_{i=0}^{D-1}$ is a set of simultaneous Schmidt decomposable
subset of canonical Bell states, then we derive
\begin{eqnarray}
A^{\dagger} \otimes A^{t} \rho A \otimes A^{*} = ( \sum _{ij}
a_{ij} \ket{\Psi _i } \bra{\Psi _j} ) \otimes \ket{\Psi _0}
\bra{\Psi _0},
\end{eqnarray}
where $A$ is a local unitary operator defined at (\ref{repA}).
 This protocol is actually entanglement distillation protocol
deriving one e-bits for all mixed state. Moreover, in the case $a
_{ij} =\delta _{ij} / D$, since $\sum _{i=0}^{D-1} \frac{1}{D}
\ket{\Psi _i}\bra{\Psi _i}$ is a separable state, this
distillation protocol by the local unitary is optimal. Actually,
the states (\ref{maximally correlated}) belong to a class of
states called ``maximally correlated states'', and the simple
formula of distillable entanglement for maximally correlated
states has been already known \cite{R01}. However the above
protocol is deterministic and moreover unitary, this is actually
important point. Generally speaking, deterministic distillable
entanglement is strictly less than usual asymptotic one
\cite{Morikoshi}. Therefore, this is a very rare case where the
meaningful lower bound of deterministic entanglement distillation
can be derived for mixed states.

\subsection{Error correction and QKD}
As another application, we can apply our result to error
correction and quantum key distribution with the following
specific noisy channel. Now, we consider the inverse of channel
copying protocols in subsection \ref{cc}, and we derive the error
correcting protocol which corresponds to the above distillation
protocol. Consider a channel $\Lambda (\rho) = \sum _{k=1}^{N} E_k
\rho E_k^{\dagger}$ on $\Hi _1 \otimes \Hi _2$, where $E_k$
satisfies $\sum _{k=1}^N E_k^{\dagger}E_k =I$ and can be written
down as $E_k = \sum _{i=0}^{D-1} c _{ki} U_i \otimes U_i$ by a
simultaneous diagonalized subset $\{ U_i \} _{i=0}^{D-1}$ of
Generalized Pauli's Group. In particular, when a channel can be
decomposed by a set of Kraus operators which have a form $E_k= p_k
U_k \otimes U_k$, the channel is called collective noise. Such a
noize may occur , for example, in the case when we send two
potonic qubits simultaneously through optical fibre or free space
\cite{collective}. Since whole dimension of operator space $\B
(\Hi _1 \otimes \Hi _2)$ is $D^4$, these error operators have very
limited forms. However, in this case, the inverse of our copying
protocol gives one noiseless channel as follows. If the channel
$\Lambda $ satisfies the above condition, then the channel
$\Lambda $ can be written down as $\Lambda (\rho ) = \sum _{ij}
a_{ij} U _i \otimes U_i \rho U _j^{\dagger} \otimes U_j^{\dagger}
$. Then, by the inverse of channel copying operation, we have the
following relation.
\begin{eqnarray*}
A^{\dagger } [ \Lambda( A \rho A^{\dagger} )] A & = & \sum _{ij}
a_{ij} A^{\dagger} (U_i \otimes U_i) A \rho A^{\dagger }
(U_j^{\dagger} \otimes U_j^{\dagger}) A \\
& = & \sum _{ij} a _{ij} (U _i \otimes I) \rho (U _j \otimes I).
\end{eqnarray*}
Thus, using an ancilla $\sigma _0$, encoding operation $A$ and
decoding operation $A^{\dagger}$, we derive a noiseless channel in
$\Hi _2$ as follows,
\begin{eqnarray*}
\Tr_1 A^{\dagger } [ \Lambda( A  (\sigma_o \otimes \sigma) A^{\dagger} )] A
= \sigma.
\end{eqnarray*}
Similarly to the distillation case,
as is shown later,
when $a_{ij} = \delta _{ij}/D$, this error correcting protocol
attains the asymptotic optimal rate of transmitting the quantum state
through the channel $\Lambda$.
That is, the transmission rate of this protocol
is equal to the quantum capacity of this rate.

This fact can be seen by the correspondence between
quantum capacity and distillable entanglement given in \cite{BDSW}. Thus,
for generalized Pauli's channel, quantum capacity coincides with distillable
entanglement of the corresponding state, which is the state
derived as the output state when inputting a part of a maximally
entangled state, {\it i.e.} $\sum _{i=0}^{D-1} \frac{1}{D}\ket{\Psi
_i} \otimes \ket{\Psi _i} \bra{\Psi _i} \otimes \bra{\Psi _i}$.
Since our protocol is the optimal distillation
protocol for this states, this channel coding protocol is also optimal.

Next, we apply this error correcting protocol to QKD. In the
two-dimensional case, applying the above encoding and decoding
protocol for usual BB84 protocol, we derive the following
protocol. Alice randomly chooses a  basis from $\{|0\rangle|0
\rangle, |1\rangle|1 \rangle\}$ and
$\{\frac{1}{\sqrt{2}}(|0\rangle|0 \rangle +|1\rangle|1 \rangle),
\frac{1}{\sqrt{2}}(|0\rangle|0 \rangle -|1\rangle|1 \rangle)\}$.
Bob performs the measurement $\{ I \otimes |0 \rangle \langle 0|,
I \otimes |1\rangle \langle 1|\}$ or $\{ \frac{1}{2} (|0\rangle|0
\rangle +|1\rangle|1 \rangle) (\langle 0|\langle 0| +\langle
1|\langle 1 |) + \frac{1}{2} (|1\rangle|0 \rangle +|0\rangle|1
\rangle) (\langle 1|\langle 0| +\langle 0|\langle 1|), \frac{1}{2}
(|0\rangle|0 \rangle -|1\rangle|1 \rangle) (\langle 0|\langle 0|
-\langle 1|\langle 1 |) + \frac{1}{2} (|1\rangle|0 \rangle
-|0\rangle|1 \rangle) (\langle 1|\langle 0| -\langle 0|\langle 1
|) \}$ with the equal probability. When Alice uses the former
basis and Bob uses the former measurement, Bob's measured data
coincides with Alice's bit in this channel. Similarly, when Alice
uses the later basis and Bob uses the later measurement, Bob's
measured data coincides with Alice's bit in this channel.
Therefore, if noise satisfies our assumption, we can realize the
noiseless QKD by the above protocol. Even if Bob cannot perform
the later, if he can realize the two-valued Bell measurement $\{
\frac{1}{2} (|0\rangle|0 \rangle +|1\rangle|1 \rangle) (\langle
0|\langle 0| +\langle 1|\langle 1 |), \frac{1}{2} (|1\rangle|1
\rangle -|0\rangle|0 \rangle) (\langle 1|\langle 1| -\langle
0|\langle 0 |), |1\rangle|0 \rangle \langle 1|\langle 0|+
|0\rangle|1 \rangle \langle 0|\langle 1| \}$, the noiseless QKD is
available by the postselection.

\section{discussion} \label{summary}
In this paper, we focus on a set consisting of several maximally
entangled states in a prime-dimensional system. In this case, we
completely characterized locally copiablity and showed the
relationship between locally copiability and local
distinguishablity. In sections \ref{local copying} and
\ref{relation}, we proved that such a set is locally copiable, if
a\begin{flushleft}

\end{flushleft}nd only if it has a canonical Bell form
and a simultaneous Schmidt decomposable (Theorem \ref{main}).
This theorem deduces the following two conclusions.
At first, as well as the maximal size of local distinguishable sets,
the maximal size of locally copiable sets is $D$, that is, equal to the dimension of the local space.
This maximal size is the square root of the maximal size without
the LOCC restriction.
Second, as we can see in FIG.\ref{hierarchy},
when such a set is locally copiable, it is also one-way locally distinguishable,
and the opposite direction is not true.
In other words,
at least in prime-dimensional systems, local copying is more difficult than
one-way local discrimination for a set of maximally entangled states.

In the case of local discrimination, a simultaneous Schmidt
decomposable set is locally distinguishable. However, if such a
set of states does not have cannonical Bell form, the set is not
locally copiable. We can interpret the above fact as follows. A
simultaneous Schmidt decomposable set does not possess
non-locality beyond individual entanglement concerning local
discrimination. On the other hand, if such a set does not have
cannonical Bell form, such a set still has non-locality concerning
local copying. In other words, we can conclude that {\it the lack
of algebraic symmetry causes extra non-locality of a set
concerning local copying.}

Although we only treated orthogonal sets of maximally entangled states
in this paper, our result of FIG.\ref{hierarchy} also regard
as the classification of sets of Schmidt basis by their non-locality.
Therefore, in the case of a set of general entangled states,
the structure of non-locality of sets of Schmidt basis may be similar to
FIG.\ref{hierarchy}, though its possesses additional non-locality
which originates in various Schmidt coefficients.
Therefore, our result may be useful as the base for more general
discussion of non-locality problems of Schmidt basis, especially
for general discussion of the local copying problems.

In section \ref{relation}, we showed that our results and protocol
of local copying can be interpret as results of several different
and closely related quantum information processing, that is, local
copying of mixed states, entanglement distillation, channel
copying, error correction, and quantum key distribution. These
close relation with many other protocols suggests the importance
of local copying as a fundamental protocol of non-local quantum
information processing.

Finally, we should mention a remained open problem.
In this paper, we showed the necessity of the form of states (\ref{nsrep}) for LOCC copying only in prime-dimensional local systems.
However, we restrict this dimensionality only by the technical reason, and
this restriction has no physical meaning. Thus, the validity of Theorem
\ref{main} for non-prime-dimensional systems still remains as a open question.

After finishing the first draft \cite{draft}, the authors found a related paper
\cite{Nathanson} which contains
a different approach to Theorem \ref{dika}.

\section*{Acknowledgments}
The authors would like to express our gratitude to Mr. F. Anselmi, Dr. A. Chefles and Professor M.B. Plenio for their useful discussion.
They would like to thank Dr. S. Virmani for useful discussion.
They are particularly indebted to Professor M. Murao for her helpful advice, discussion and checking the introduction.
They are grateful to Professor K. Matsumoto for his advice, Dr. D. Markham for discussion, and
Professor H. Imai for his support and encouragement.
They are also grateful Professor M.B. Ruskai for informing them the paper \cite{Nathanson}.


\appendix
\section{Proof of Lemma \ref{lemma}}
First, by choosing $c = a_1 \oplus b_1 = a_2 \oplus b_2$, we have
\begin{equation} \delta _{b_1 b_2} U_{cc} = U_{c \ominus b_1 \ c \ominus b_2} U_{b_1 b_2}.
\end{equation}
In addition, choosing $b_1 \neq b_2$, we derive
\begin{equation}
U_{c \ominus b_1 \ c \ominus b_2} U_{b_1 b_2} =0
\end{equation}
for all $c$.
The above equation means,
\begin{equation} \label{eqlemma2}
 b_1 \neq b_2 \quad \Longrightarrow \quad U_{b_1 b_2} = 0 \quad or \quad \forall c, \  U_{c \ominus b_1 \ c \ominus b_2} = 0.
\end{equation}

By means of the above fact, we prove that $U_{b \ b \oplus n} = U_{b \ b \ominus n}=0$ for all $b$ by
induction concerning the integer $n$.
At first, we prove  $U_{b \ b \ominus 1} = 0$ for all $b$ by a contradiction.
We assume there exists $b_1$ such that $U_{b_1 \ b_1 \ominus 1} \neq 0$,
then, Equation (\ref{eqlemma2}) implies  $U_{b \ b+1} = 0$ for all $b$.
On the other hand, Equation (\ref{eqlemma1}) guarantees that
\begin{equation} \label{keylemma}
\Xi _{b_1 \ b_1 \ominus 1}^{b \ b+1} U_{b \ b+1} = U_{b \ominus b_1 \ b \ominus b_1 \oplus 2} U_{b_1 \ b_1 \ominus 1}.
\end{equation}
Therefore,  $U_{b \ b \oplus 2} = 0$ for all $b$. Thus, repeatedly
using (\ref{eqlemma1}), we have $U_{b \ b \oplus a} = 0$ for all
$a$ and $b$. This is a contradiction for the fact that $U_{ab}$ is
a non-zero matrix. So, we have $U_{b \ b \ominus 1} = 0$ for all
$b$. Similarly, we can prove $U_{b \ b \oplus 1} =0$ for all $b$
by a contradiction as follows. Suppose there exists $b_1$ such
that $U_{b_1 \ b_1 \oplus 1}=0$, then Equation (\ref{eqlemma1})
implies $\Xi _{b_1 \ b_1 \ominus 1}^{b \ b \ominus 1} U_{b \ b
\ominus 1} = U_{b \ominus b_1 \ b \ominus (b_1 - 2)} U_{b_1 \ b_1
\oplus 1}$. Therefore, $U_{b \ b \ominus 2} =0$ for all $b$, and
repeating this procedure, we have $U_{b \ b \ominus a} =0$ for all
$a$ and $b$. This is a contradiction. Therefore,  $U_{b \ b \oplus
1} = 0$ for all $b$.

At the next step, we assume  $U_{b \ b \ominus k} = U_{b \ b \oplus k} = 0$ for all $k \le n-1$ and show
$U_{b \ b \ominus n} = U_{b \ b \oplus n} = 0$ for any $b$ by a contradiction.
Assume that there exists $b_1$ such that  $U_{b_1 \ b_1 \ominus n} \neq 0$, then Equation (\ref{keylemma}) implies that $U _{b \ b \oplus n} = 0$ for all $b$.
Thus, Equation (\ref{eqlemma1}) implies
\begin{equation}
\Xi _{b_1 \ b_1 \ominus 1}^{b \ b \oplus k} U_{b \ b \oplus k} = U_{b \ominus b_1 \ b \ominus b_1 \oplus (n + k)} U_{b_1 \ b_1 \ominus n}.
\end{equation}
Thus, we have $U_{b \ b \oplus (n+k)} =0$ for all $k \le n-1$ and $b$.
Repeating this procedure, we have $U_{b \ b \oplus a} = 0$ for all $a$ and $b$.
This is a contradiction.
Therefore, we have $U_{b \ b \ominus n} = 0$ for all $b$.
Similarly, we can prove $U_{b \ b \oplus n} = 0$ for all $b$.
Finally, by the mathematical induction, we prove $U_{b \ b \oplus n} = U _{b \ b \ominus n} = 0$
for all $0 \le n < D$ and  $b$.
Therefore, $U_{ab}$ is a diagonal matrix.
\hfill $\square$


\end{document}